\documentclass[pra,twocolumn]{revtex4}%
\usepackage{amsfonts}
\usepackage{amsmath}
\usepackage{amssymb}
\usepackage{subfigure}
\usepackage{graphicx}%
\providecommand{\U}[1]{\protect\rule{.1in}{.1in}}

\begin{document}
\title{Cavity linewidth narrowing with dark-state polaritons }
\author{G. W. Lin$^{1}$}
\email{gwlin@ecust.edu.cn}
\author{J. Yang$^{1}$}
\author{X. M. Lin$^{2}$}
\author{Y. P. Niu$^{1}$}
\email{niuyp@ecust.edu.cn}
\author{S. Q. Gong$^{1}$}
\email{sqgong@ecust.edu.cn}
\affiliation{$^{1}$Department of Physics, East China University of Science and Technology,
Shanghai 200237, China}
\affiliation{$^{2}$School of Physics and Optoelectronics Technology, Fujian Normal
University, Fuzhou 350007, China}

\begin{abstract}
We perform a quantum-theoretical treatment of cavity linewidth narrowing with
intracavity electromagnetically induced transparency (EIT). By means of
intracavity EIT, the photons in the cavity are in the form of cavity
polaritons: bright-state polariton and dark-state polariton. Strong coupling
of the bright-state polariton to the excited state induces an effect known as
vacuum Rabi splitting, whereas the dark-state polariton decoupled from the
excited state induce a narrow cavity transmission window. Our analysis would
provide a quantum theory of linewidth narrowing with a quantum field pulse at
the single-photon level.

\end{abstract}

\maketitle

\section{\textbf{Introduction}}

Electromagnetically induced transparency (EIT) can be used to make a resonant,
opaque medium transparent by means of a strong coupling field acting on the
linked transition \cite{Harris,Fleischhauer}. The EIT medium in an optical
cavity known as intracavity EIT was first discussed by Lukin et al.
\cite{Lukin}. They show that the cavity response is drastically modified by
intracavity EIT, resulting in frequency pulling and a substantial narrowing of
spectral features. By following this seminal work, significant experimental
advance has been made in narrowing the cavity linewidth
\cite{Wang,Wu,Hernandez,Wu1} and enhancing the cavity lifetime \cite{Laupr}.

The previous semi-classical treatments of cavity linewidth narrowing and
lifetime enhancing with intracavity EIT were based on the solution of the
susceptibility of EIT system. Here we present a quantum-theoretical treatment
of cavity linewidth narrowing with intracavity EIT. By means of intracavity
EIT, the photons in the cavity are in the form of cavity polaritons:
bright-state polariton and dark-state polariton. Strong coupling of the
bright-state polariton to the excited state induces an effect known as vacuum
Rabi splitting, whereas the dark-state polariton decoupled from the excited
state induces a narrow cavity transmission window $\upsilon=\cos^{2}%
\theta\upsilon_{0}$, with $\upsilon_{0}$ the empty-cavity linewidth and
$\theta$ the mixing angle of the dark-state polariton \cite{Fleischhauer1}. We
discuss the condition required for cavity linewidth narrowing and find that
when the atom-cavity system is in the collective strong coupling regime, a
weak control field is sufficient for avoiding the absorption owing to
spontaneous emission of the excited state, and then the dark-state polariton
induces a very narrow cavity linewidth. This result is different from that
based on previous semi-classical treatments of intracavity EIT
\cite{Lukin,Wang}, where the strong control field is required for avoiding the
absorption of a probe field pulse if all atoms are prepared in one of their
ground states. \begin{figure}[ptb]
\includegraphics[width=3.2in]{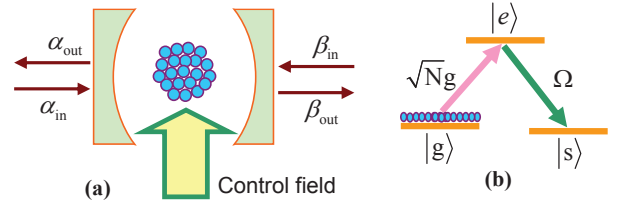}\newline\caption{ (Color online) (a)
Schematic setup to cavity linewidth narrowing with dark-state polaritons. (b)
The relevant atomic level structure and transitions.}%
\end{figure}

\bigskip

\section{Semi-classical theory}

We first review the intracavity EIT with the semi-classical theory based on
the solution of the susceptibility of EIT system. An EIT medium of length $l$
is trapped in an optical cavity of length $L$. The EIT medium response of a
probe classical field is characterized by the real $\chi^{^{\prime}}$ and the
imaginary $\chi^{^{\prime\prime}}$ parts of the susceptibility. The real part
gives the dispersion $\frac{\partial\chi^{^{\prime}}}{\partial\omega_{p}}$ and
the imaginary part gives the absorption coefficient $\alpha=2\pi\omega_{p}%
\chi^{^{\prime\prime}}/c$, with probe frequency $\omega_{p}$. Then the ratio
of the linewidth $\upsilon$ of the cavity with the EIT medium to that of the
empty cavity is \cite{Lukin,Wang}%

\begin{equation}
\frac{\upsilon}{\upsilon_{0}}=\frac{1-r\tau}{\sqrt{\tau}(1-r)}\frac{1}{1+\eta
},
\end{equation}
where $\tau=\exp(-\alpha l)$, $\eta=\omega_{r}(l/2L)\frac{\partial
\chi^{^{\prime}}}{\partial\omega_{p}}$, with $\omega_{r}$ the cavity resonant
frequency, and $r$ is the intensity reflectivity of the cavity mirror. As
shown in Ref \cite{Lukin,Wang}, when the EIT medium is driven by a strong
control field, the absorption of the probe field can be negligible
($\chi^{^{\prime\prime}}\rightarrow0$), whereas the dispersion is large,
resulting in a substantial narrowing of the cavity linewidth.

\section{\textbf{Quantum-theoretical treatment}}

Now we try to solve the quantum dynamics of EIT in an optical standing-wave
cavity [Fig, 1(a)]. We consider an atomic system with three levels, two ground
states $\left\vert g\right\rangle $, $\left\vert s\right\rangle $, and an
excited $\left\vert e\right\rangle $, forming a $\Lambda$-configuration [see
Fig, 1(b)]. The two transitions $\left\vert g\right\rangle \leftrightarrow
\left\vert e\right\rangle $ and $\left\vert s\right\rangle \leftrightarrow
\left\vert e\right\rangle $ are resonantly coupled by a cavity mode and a
laser fields, respectively. The interaction Hamiltonian for the coherent
processes is described by $H_{I}=%
{\displaystyle\sum\nolimits_{j=1}^{N}}
(g\left\vert e\right\rangle _{j}\left\langle g\right\vert a+\Omega\left\vert
s\right\rangle _{j}\left\langle e\right\vert +H.c.)$, where $a$ is the
annihilation operator of the cavity mode, $g$ ($\Omega$) is the coupling
strength of quantized cavity mode (external field) to the corresponding
transition. We assume that almost all atoms are in one of their ground states,
e.g. $\left\vert G\right\rangle =\prod_{j=1}^{N}\left\vert g_{j}\right\rangle
$, at all times, and define the collective atomic operators $C_{\mu}^{\dagger
}=\frac{1}{\sqrt{N}}%
{\displaystyle\sum\nolimits_{j=1}^{N}}
\left\vert \mu\right\rangle _{j}\left\langle g\right\vert $ with $\mu=e,s$,
then $H_{I}$ can be rewritten as%
\begin{equation}
H_{I}^{^{\prime}}=\sqrt{N}gC_{e}^{\dagger}a+C_{s}^{\dagger}C_{e}\Omega+H.c.,
\end{equation}
where the coupling constant $g$ between the atoms and the quantized cavity
mode is collectively enhanced by a factor $\sqrt{N}$. In analogy to the
dark-state polariton for a travelling light pulse \cite{Fleischhauer1}, one
can define two standing-wave cavity polaritons: a dark-state polariton
$m_{D}=\cos\theta a-\sin\theta C_{s}$, and a bright-state polariton
$m_{B}=\sin\theta a+\cos\theta C_{s}$, with $\cos\theta=\Omega_{c}%
/\sqrt{Ng^{2}+\Omega_{c}^{2}}$ and $\sin\theta=\sqrt{N}g/\sqrt{Ng^{2}%
+\Omega_{c}^{2}}$. In terms of these cavity polaritons the Hamiltonian
$H_{I}^{^{\prime}}$ can be represented by%

\begin{equation}
H_{I}^{^{^{\prime\prime}}}=\sqrt{Ng^{2}+\Omega^{2}}(C_{e}^{\dagger}m_{B}%
+C_{e}m_{B}^{\dagger}).
\end{equation}
The cavity dark-state polariton is decoupled from the collective excited state
$C_{e}^{\dagger}\left\vert G\right\rangle $.

The external fields interact with cavity mode $a$ through two input ports
$\alpha_{in}$, $\beta_{in}$, and two output ports $\alpha_{out}$, $\beta
_{out}$.\ The Hamiltonian for the cavity input--output processes is described
by \cite{Walls} $H_{in-out}=%
{\displaystyle\sum\nolimits_{\Theta=\alpha,\beta}}
{\displaystyle\int\nolimits_{-\infty}^{+\infty}}
\omega d\omega\Theta^{\dagger}(\omega)\Theta(\omega)+%
{\displaystyle\sum\nolimits_{\Theta=\alpha,\beta}}
[i%
{\displaystyle\int\nolimits_{-\infty}^{+\infty}}
d\omega\sqrt{\frac{\kappa}{2\pi}}\Theta^{\dagger}(\omega)a+H.c.],$ where
$\omega$ is the frequency of the external field, $\kappa$ is the bare cavity
decay rate without EIT medium, and $\Theta(\omega)$ with the standard relation
$[\Theta(\omega),\Theta^{\dagger}(\omega^{^{\prime}})]=\delta(\omega
-\omega^{^{\prime}})$ denotes the one-dimensional free-space mode. We express
the Hamiltonian $H_{in-out}$ in the polariton bases:%

\begin{align}
H_{in-out}^{^{\prime}}  &  =%
{\displaystyle\sum\limits_{\Theta=\alpha,\beta}}
{\displaystyle\int\nolimits_{-\infty}^{+\infty}}
\omega d\omega\Theta^{\dagger}(\omega)\Theta(\omega)\nonumber\\
&  +%
{\displaystyle\sum\limits_{\Theta=\alpha,\beta}}
[i%
{\displaystyle\int\nolimits_{-\infty}^{+\infty}}
d\omega\Theta^{\dagger}(\omega)(\sqrt{\frac{\kappa_{_{D}}}{2\pi}}%
m_{D}\nonumber\\
&  +\sqrt{\frac{\kappa_{_{B}}}{2\pi}}m_{B})+H.c.],
\end{align}
with $\kappa_{_{D}}=\cos^{2}\theta\kappa$ and $\kappa_{_{B}}=\sin^{2}%
\theta\kappa$.

In the intracavity EIT system, only the bright-state polariton $m_{B}$
resonantly couples to the excited state. Under the condition
\begin{equation}
\sqrt{Ng^{2}+\Omega^{2}}\gg\kappa_{_{B}},\gamma_{e},
\end{equation}
with $\gamma_{e}$ the spontaneous-emission rate of the excited state
$\left\vert e\right\rangle $, the resonant interaction in Hamiltonian
$H_{I}^{^{^{\prime\prime}}}$ in the so-called strong coupling regime will
induce an effect known as vacuum Rabi splitting \cite{Boca}, i.e., the
splitting of the cavity transmission peak for the bright-state polariton
$m_{B}$ into a pair of resolvable peaks at $\omega=\omega_{0}\pm\sqrt
{Ng^{2}+\Omega^{2}}$ (here $\omega_{0}$ is the resonant frequency of cavity
mode). Thus one can neglect bright-state polariton $m_{B}$ to calculate the
cavity transmission spectrum near the cavity resonant frequency $\omega_{0}$.

According to quantum Langevin equation, the evolution equation of the cavity
dark-state polariton $m_{D}$ is given by \cite{Walls}\begin{figure}[ptb]
\includegraphics[width=2.6in]{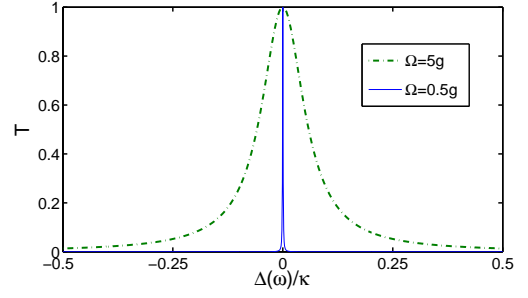}\newline\caption{(Color online) The
transmission spectrum $T$ as a function of $\Delta(\omega)$ for $\Omega
=\{5g,0.5g\}$, on the assumption that $N=400$, $g=\kappa=\gamma_{e}.$ }%
\end{figure}%
\begin{equation}
\dot{m}_{D}=-i\omega_{0}m_{D}-\kappa_{_{D}}m_{D}+\sqrt{\kappa_{_{D}}}%
\alpha_{in}+\sqrt{\kappa_{_{D}}}\beta_{in}.
\end{equation}
Using the relationships between the input and output modes at each mirror
\cite{Walls}%

\begin{equation}
\alpha_{out}\left(  t\right)  +\alpha_{in}\left(  t\right)  =\sqrt
{\kappa_{_{D}}}m_{D},
\end{equation}
and%
\begin{equation}
\beta_{out}\left(  t\right)  +\beta_{in}\left(  t\right)  =\sqrt{\kappa_{_{D}%
}}m_{D},
\end{equation}
and the Fourier transformations: $\Lambda=\sqrt{\frac{1}{2\pi}}%
{\displaystyle\int\nolimits_{-\infty}^{+\infty}}
d\omega\Lambda(\omega)e^{-i\omega t}$, with $\Lambda=m_{D},$ $\alpha
_{in},\alpha_{out},\beta_{in},\beta_{out}$, we can find
\begin{equation}
\alpha_{out}\left(  \omega\right)  =\frac{\kappa_{_{D}}\beta_{in}\left(
\omega\right)  }{\kappa_{_{D}}-i\Delta(\omega)},
\end{equation}
where $\Delta(\omega)=\omega-\omega_{0}$, and we have assumed that the photons
enter into the cavity from the input port $\beta_{in}$ ($\alpha_{in}=0$). Then
the transmission spectrum for intracavity EIT is described by%

\begin{equation}
T(\omega)=\frac{\left\vert \alpha_{out}\left(  \omega\right)  \right\vert
^{2}}{\left\vert \beta_{in}\left(  \omega\right)  \right\vert ^{2}}%
=\frac{\kappa_{_{D}}^{2}}{\kappa_{_{D}}^{2}+\Delta^{2}(\omega)}.
\end{equation}
As depicted in Fig. 2, the transmission spectrum $T$ can be controlled by the
external coherent field. Then the calculate of cavity linewidth $\Delta
_{\upsilon}$, i.e., the full width at half height of $T(\omega)$, gives%

\begin{equation}
\upsilon=2\kappa_{_{D}}=2\kappa\cos^{2}\theta=\cos^{2}\theta\upsilon_{0},
\end{equation}
here $\upsilon_{0}=2\kappa$ is the empty-cavity linewidth \cite{Walls}.

\section{\textbf{Brief discussion }}

Next we briefly discuss the results of the quantum-theoretical treatment of
intracavity EIT. First, the polariton $m_{D}$ corresponds to the well-known
\textquotedblleft dark-state polariton\textquotedblright\ for a travelling
light pulse \cite{Fleischhauer1}. In Ref \cite{Fleischhauer1}, the mixing
angle $\theta$ determines the group velocity of dark-state polariton, whereas
the mixing angle $\theta$ of cavity dark-state polariton $m_{D}$ here
determines the effective cavity decay rate $\kappa_{_{D}}$. Second, the atomic
ground states have long coherence time and very small decay rate, thus the
main source of absorption by EIT system is spontaneous emission of the excited
state. To avoid the absorption of the probe field by the coupling of
bright-state polariton $m_{B}$ to excited state, we require the condition in
Eq. (5). If atom-cavity system is in the weak coupling regime $\sqrt
{N}g<\kappa,\gamma_{e}$, we need a strong control field, $\Omega\gg g$, to
satisfy the required condition. However, if atom-cavity system is in the
collective strong coupling regime $\sqrt{N}g\gg\kappa,\gamma_{e}$, even when
the control field is so weak that $\Omega\ll g$, the required condition is
still satisfied and the absorption owing to spontaneous emission of the
excited state can be neglected, then the dark-state polariton will induce a
very narrow cavity linewidth $\upsilon=$ $\cos^{2}\theta\upsilon_{0}%
\approx\upsilon_{0}\Omega^{2}/Ng^{2}$. We note this result is different from
that based on previous semi-classical treatments of intracavity EIT
\cite{Lukin,Wang}, where a strong control field is required for avoiding the
absorption of a probe field pulse if initially (i.e., before the probe field
arrives) all atoms are in their ground states $\left\vert G\right\rangle
=\prod_{j=1}^{N}\left\vert g_{j}\right\rangle $.

\section{\textbf{Conclusion }}

In summary, we have performed a theoretical investigation of intracavity EIT
quantum mechanically. In intracavity EIT system the cavity photons are in the
form of cavity polaritons: bright-state polariton and dark-state polariton.
Strong coupling of the bright-state polariton to the excited state leads to an
effect known as vacuum Rabi splitting, whereas the dark-state polariton
decoupled from the excited state induce a narrow transmission window. If
atom-cavity system is in the weak coupling regime, a strong control field is
requied for avoiding the absorption owing to spontaneous emission of the
excited state. However, if atom-cavity system is in the collective strong
coupling regime, a weak control field is sufficient for avoiding the
absorption, and then the dark-state polariton induces a very narrow cavity
linewidth. This result is different from that based on previous semi-classical
treatments of intracavity EIT \cite{Lukin,Wang}, where the strong control
field is required for avoiding the absorption of a probe field pulse if all
atoms are prepared in one of their ground states. Our quantum-theoretical
treatment of intracavity EIT would provide a quantum theory of linewidth
narrowing with quantum field pulses at the single-photon level.

\textbf{Acknowledgments: }This work was supported by the National Natural
Sciences Foundation of China (Grants No. 11204080, No. 11274112, No. 91321101,
and No. 61275215), the Fundamental Research Funds for the Central Universities
(Grants No. WM1313003).

\end{document}